# Monitoring of Real Time ECG Signals on Mobile System

1st Beyazit B Yuksel
*Computer Engineering*
*Yıldız Technical University*
Istanbul, Turkey
yukselbestami@hotmail.com

2nd Paşa Yazıcı
*Computer Engineering*
*Yıldız Technical University*
Istanbul, Turkey
pasayazici@hotmail.com

3rd Gökhan Bilgin
*Computer Engineering*
*Yıldız Technical University*
Istanbul, Turkey
gokhanb@yildiz.edu.tr

*Abstract—* This study focuses on the connection of a development kit that enables real-time monitoring of electrocardiogram (ECG) signals using a mobile system. A software developed on the Visual Studio .NET platform reads real-time ECG signals from the human body through non-invasive methods and displays them graphically on the mobile system. ECG electrodes placed on specific areas of the body using the method known as Einthoven's triangle. Subsequently, the software initiates data flow through the serial port, and these data displayed as signal values on the mobile device's screen via a graphical interface. When the monitored ECG signals fall below a certain threshold or reach a critical value, the system provides feedback with an alert based on medical data. The developed system is fully portable. Additionally, the implemented system has the potential to form the basis for a multi-purpose system in the future, such as online patient monitoring, patient location tracking, and even initial intervention using the defibrillation method.

## I. INTRODUCTION

Real-time signal monitoring plays a crucial role in the field of biomedical engineering. In this study, a system was initially developed on the .NET platform to provide a software-based infrastructure for processing biomedical signals related to the heart. This system includes the integration of an ECG (electrocardiography) preamplifier module, which allows the acquisition of biological heart signals from the human body through non-invasive methods. Subsequently, a software was created to enable continuous graphical monitoring of the information obtained from the ECG module, featuring an interactive interface. The system also measures pulse rates and issues an alert when the values fall below a certain threshold or reach a critical level, based on medical pre-information. The second part of the system is the mobile software. Here, the signal values obtained from the body are sent to a server via a web service, and real-time or historical records can be monitored on mobile platforms equipped with the implemented application. Therefore, the developed system enables remote patient monitoring.

## II. ELECTROCARDIOGRAPHIC SIGNALS

In an ECG, each heartbeat corresponds to a signal complex consisting of waves named P, QRS, and T, along with straight lines between them [1]. This signal is illustrated in Figure 1. The accurate interpretation of ECG signals primarily depends on the quality of the ECG recordings obtained from the patient [12]. Therefore, while measuring ECG signals from the patient, the patient should be informed about the procedure, and electrodes, which convert ion currents to electron currents, should be placed in the correct positions on the skin with proper contact after applying a sufficient amount of electro-gel. To protect the ECG signals from disturbances and noise that may distort the structure, proper grounding is essential. Taking this into consideration, the electrodes should be attached to the patient in the correct sequence [2, 4].

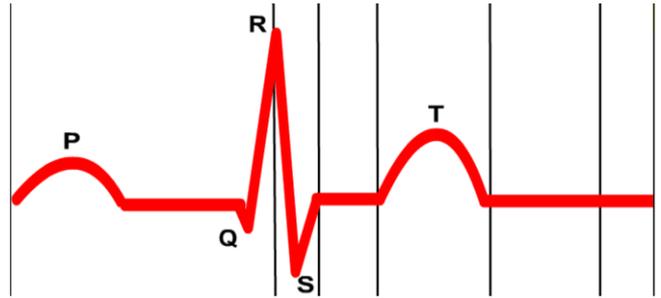

*Figure 1: ECG Signal Complex*

The formation of the ECG signal complex occurs through the following steps:

1. The SA node is triggered.

2. A very small electrical potential is generated in the atrial muscle tissue, causing atrial contraction. This results in the formation of the P wave in the ECG. Due to the slow contraction of the atria, the P wave is a small wave, with an amplitude of 0.1 to 0.2 mV and a duration of 60 to 80 ms.

3. The excitation wave pauses at the AV node, creating a delay of approximately 60 to 80 ms after the P wave. This interval is known as the PQ interval, during which blood flows from the atria to the ventricles.

4. The bundle of His, bundle branches, and Purkinje fibers facilitate rapid excitation of the atria.

5. The excitation wave propagates from top to bottom, causing rapid ventricular contraction. This forms the

QRS wave in the ECG, characterized by an amplitude of approximately 1 mV and a duration of around 80 ms. It is a sharp wave.

6. Following ventricular contraction, the ST interval, which lasts for about 100 to 120 ms, is observed.

7. The relaxation of the ventricles results in the T wave. The T wave has an amplitude of 0.1 to 0.3 mV and a duration of 20 to 160 ms [6, 8].

### A. Three-Lead ECG

The electrical activity of the heart can be represented as a dipole, meaning it acts as a vector between two poles of nerve cells. The theory that explains the electrical events observed during the excitation of the heart muscle is known as the dipole theory [6, 10]. The positions of the electrodes placed on the body determine a vector configuration as a function of time. Figure 3 illustrates the most basic attachment methods of electrodes to the body according to Einthoven's triangle. Each corner of this theoretical triangle, marked as L1, L2, and L3 around the heart, represents the locations where the heart is electrically connected to surrounding fluids and other organs. Lead I shows the potential difference between the right and left arms, Lead II shows the potential difference between the right arm and the left leg, and Lead III shows the potential difference between the left arm and the left leg. According to Einthoven's law, if two leads are known, the third lead can be calculated. This principle is crucial in the implementation of the design as it simplifies the detailed design and reduces the number of system components to just two differential amplifiers, as required by the system [3, 9].

### III. PRELIMINARY

The diagram illustrating the system's operation is shown in Figure 2. In this setup, the signals can be monitored synchronously across the mobile systems and other client system's used.

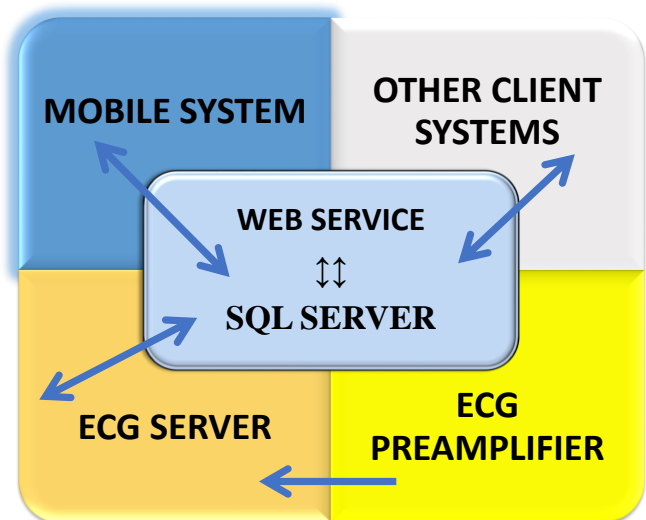

*Figure 2: System Overview*

Figure 4 provides a general overview of the components that constitute the system. In this configuration, the data read from the serial port is used to display real-time signals, and simultaneously, this data is also transmitted to the server.

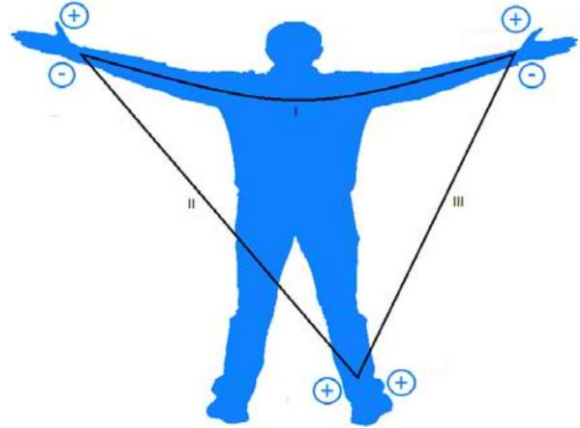

*Figure 3: Three-Lead ECG Connection and Einthoven's Triangle*

A USB-TTL adapter used to connect the ECG development kit shown in Figure 6 to the server. A three-lead ECG cable connected to the other three inputs. The voltage required for the system to operate is 5V. The connections between the module, probes, and the USB-TTL adapter illustrated in Figure 5.

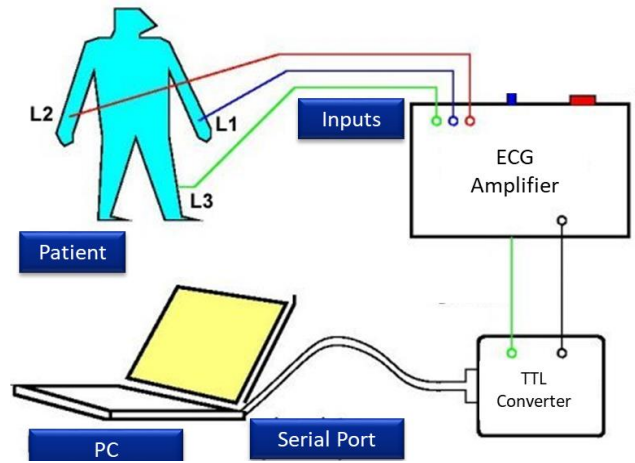

*Figure 4: General Structure of Realized System*

At the very beginning of the system, a high-pass filter with a cutoff frequency of 0.05 Hz used to suppress low-frequency noise. At the end of the high-pass filter, a low-pass filter with a cut-off frequency of 150 Hz employed. Suppressing noise from the city power grid and high-frequency noise is particularly important in this context. The output of the low-pass filter converted into a digital value detectable by the computer using an ADC. To connect the ECG circuit to the computer, an optical isolation circuit with a high insulation resistance is used [11-15].

*a) EG01000 Development Module:* The system discussed in this paper utilizes the ECG development kit (Figure 6-7). The EG01000 is a module that measures a patient's ECG in a single channel through a three-lead cable. This ECG circuit measures the voltage values produced by the heart. The voltage values of the signals generated by the heart range from 0.2 mV to 2 mV, and the bandwidth varies between 0.05 and 150 Hz. An OP-AMP with high CMRR and gain values is used at the input [7]. The value measured by the system is the differential input voltage between the red and yellow leads (clips). These clips carry the differential voltage to the input of the amplifier. The third cable, which is typically colored green, serves as the ground connection for the module.

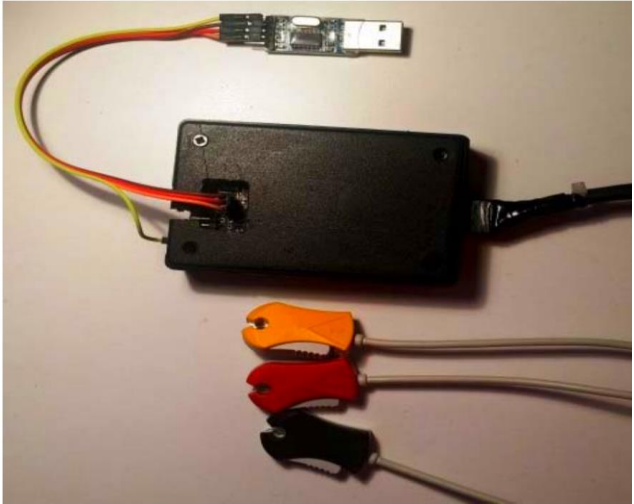

*Figure 5: Connection of the Probes to the Module and Adapter*

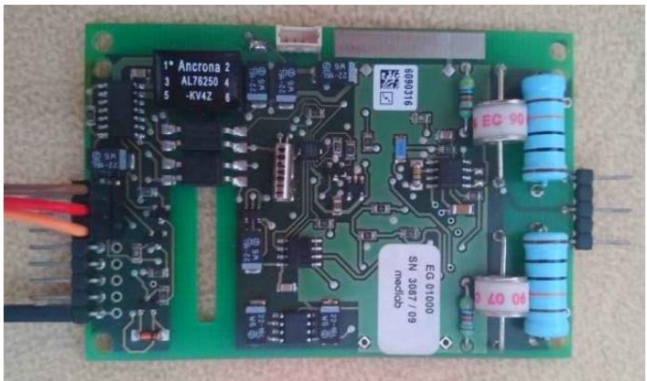

*Figure 6: EG01000 Development Module*

*b) Software Design:* The interface of the developed software, shown in Figure 7, was created on the Visual Studio .NET 2015 platform. This software features an interactive interface. Information related to the doctor who logs into the system and the registered patients is loaded into the form. The data read from the serial port is interpreted based on the information provided in Table 1 and displayed on the screen as output.

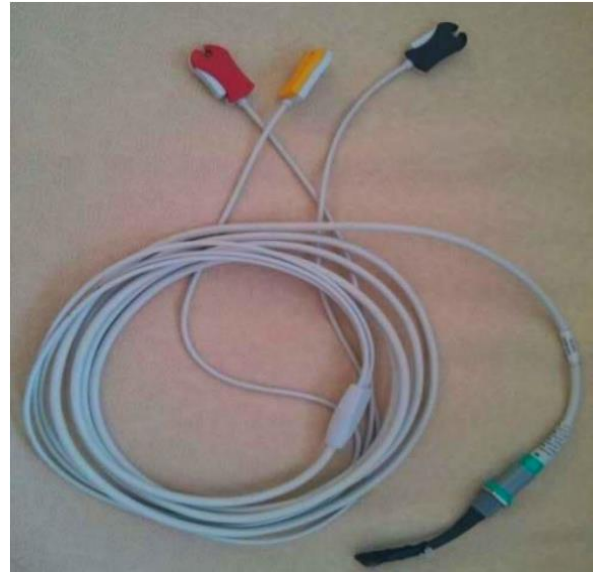

*Figure 8: Three Lead-ECG Cable*

*c) Serial Communication:* The scaling of ECG curves is done in physical units of "s/mm" for the x-axis and "mV/cm" for the y-axis. In the EG01000, data flow and command reception are continuous. Commands are received as single-byte characters. According to the data flow example shown in Table 1, the received data is interpreted and plotted on the screen. All data is transmitted at 9600 baud, 8-bit, 1 stop bit, with no parity bit. The average pulse rate calculated for each detected heartbeat, as determined by the module's internal algorithm, is sent to the port as a data block. The sampling points of the ECG curves range from decimal values 0 to 246. Sampling points with values higher than 246 (0xF6) are marked as new data values according to the definitions provided in the following Table 2.

TABLE I. THE ECG WAVE SAMPLE POINTS

| Marker byte | Meaning of following byte(s) |
|---|---|
| 0xF8 | wave sample points follow |
| 0xFA | Pulse value follows |
| 0xFB | Info byte follows |
| 0x11 | The only info byte defined is 0x11, "LEAD OFF" |

TABLE II. EXAMPLE OF A RECEIVED DATA STREAM AT THE HOST SIDE

| | Data Stream | | | | |
|---|---|---|---|---|---|
| **Byte** | 0xF8 | 0x20 0x23 0x25 | 0xFA 0x80 | 0xF8 | 0x24 0x25 0x26 |
| **Mean** | Wave Marker | ECG sample points | Pulse = 128 | Wave Marker | ECG Sample points |
| **Time** | --------------------------------→ | | | | |

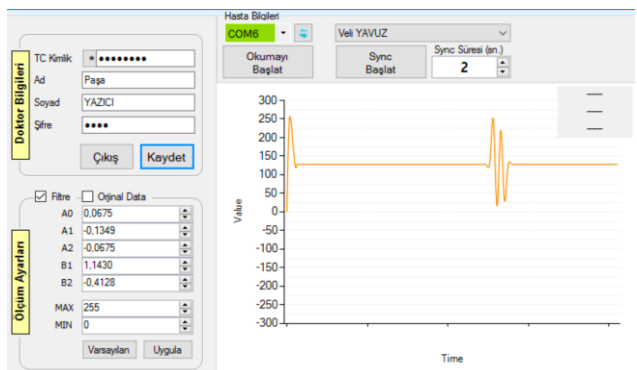

*Figure 7: Connection of the Probes to the Module and Adapter*

*d) Filtering:* Due to the presence of some high-frequency noise signals in the output from the ECG preamplifier circuit, an additional software-based high-pass filter was designed. MATLAB's filter toolbox was used for this purpose. After various trials, a 4th-order Butterworth filter was determined to be the most suitable option. The Butter function used provides the transfer function for the digital filter as shown in Equation (1) below, given the desired cutoff frequency (150 Hz) and sampling frequency (300 Hz) [13].

$$H(z) = \frac{0.0675 - 0.1349\,z^{-1} - 0.0675\,z^{-2}}{1 + 1.143\,z^{-1} - 0.4128\,z^{-2}} \quad (1)$$

## IV. Experimental results

In this section, samples taken from patients of different age groups and genders presented as shown in Figures 9, 10, and 11. The samples obtained from the mobile application both with and without filtering (Figures 9-10). The real-time samples can monitored synchronously from another remote client system. Reversing the probe connections on the wrists causes a change in the direction of the signal. Additionally, in desktop and web applications, the patient's pulse rate can also be displayed. The real-time signals can monitored simultaneously through the web server and synchronized mobile systems after the patient and doctor have logged in, and it is fully portable due to its built-in battery.

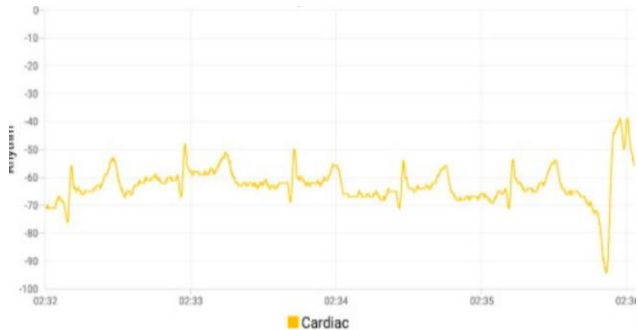
*Figure 9: ECG Sample -1*

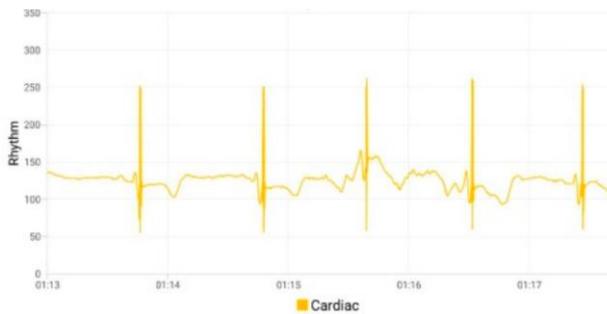
*Figure 10: ECG Sample -2*

## V. Conclusions and Future Works

In this study, it has been challenging to obtain a flawless signal due to external factors such as noise, ambient temperature, and the lack of complete isolation. However, the system has been reviewed by a specialist cardiologist, who confirmed the accuracy of the data provided and validated that the system can be used for its intended purpose. The data obtained from the ECG amplifier do not contain time information. Therefore, the acquired data were processed through a mathematical function to generate time information and buffered along with the data read from the port. A literature review has shown that studies conducted on the development of PDA-based real-time ECG monitoring systems and software for remotely controlled heart devices. In this system, however, the software implemented on mobile platforms allows the system to determine the patient's location using GPS and send the necessary data to the relevant person. This makes it easier to reach a specialist doctor. The system also has the potential to provide the infrastructure for sudden interventions such as defibrillation.

## Acknowledgement

This study presented on May 16, 2017, at the 2nd National Symposium on Biomedical Device Design and Manufacturing (UBİCTÜS), organized by the Biomedical Electronics Design, Application, and Research Center of Fatih Sultan Mehmet Vakif University.